\documentclass[amsmath,amssymb,lengthcheck,aps,prl] {revtex4}
\usepackage[colorlinks,allcolors=blue]{hyperref}
\usepackage[T1]{fontenc}
\usepackage[utf8]{inputenc}
\usepackage{lmodern}
\usepackage{graphics}
\usepackage{graphicx}

\begin{document}

\title{Study of partially polarized fractional quantum Hall states}
\author{Moumita Indra, Debashis Das, Dwipesh Majumder}
\affiliation{Department of Physics, Indian Institute of Engineering Science and Technology, Howrah, W.B., India}

\begin{abstract}

We have studied the partially spin polarized fractional quantum Hall states using Chern Simon's theory and plasma picture proposed by Halperin. Using these theoretical techniques we have tried to find the stable polarized states of different filling fractions observed in experiments. We have calculated the ground state energies of those states and also pair correlation function. We have described the nature of the states by the behavior of this quantity. In our study, we have seen that the partially polarized states, which do not fit with Jain's composite fermion description are basically the mixed state of up-spin liquid phase and down-spin solid phase.

\end{abstract}

\maketitle

The fractional quantum Hall effect (FQHE)\cite{fqhe} is basically the problem of interacting electrons in two-dimensional electron gas (2DEG) in the presence of a strong perpendicular magnetic field. The well established Composite Fermion (CF) theory\cite{Jain_CF, Jain_CF1} is based on the principle that, in a range of filling factors, each electron in the lowest Landau level (LL) captures an even number of quantum mechanical vortices to the many-particle wave function. The bound state of an electron and vortices behaves the same as a single particle, called the composite fermion\cite{Jain_CF}, which experiences a reduced amount of magnetic field $ B^*= B\pm 2p\rho \phi_0 $, where  $2p$  is an even integer number of flux attachment with each electron,  $B$ is external magnetic field, $ \rho $ is the electron (CF) density, and $ \phi_0$ is the flux quantum. CF's form their own Landau-like kinetic energy levels in this reduced magnetic field, called $ \Lambda $ levels, and their filling factor $ \nu^{*} $ is related to the electron filling factor $ \nu $ through the relation, $\nu =\frac{\nu^{*} }{2p\nu ^{*}\pm 1}$. In particular, at $ \nu =\frac{n}{2pn\pm 1} $, the ground state consists of n filled $ \Lambda $ levels. CF theory explains the FQHE states in details qualitatively. Collective excitation of almost all the filling fraction in the Jain series of positive flux attachment states has been studied earlier\cite{DDMI, DM_SSM}. 

In strong magnetic field, spin degree of freedom of electron gets frozen in the direction of the magnetic field. Thus, we obtain fully polarized quantum Hall (QH) states. Partially polarized QH states are found in the experiments\cite{a,b,c,d} for relatively small tilted magnetic field, in which the Zeeman splitting energy is small compared to cyclotron energy. Landau level mixing plays an important role in spin polarized FQHS\cite{LLMixing}, which breaks the particle-hole symmetry. Exact diagonalization\cite{exact_dia} method has been used to study the partially polarized states using small number of particles with thermodynamic extrapolation\cite{extrapolation}. S. Mandal and Ravishankar proposed a global doublet model\cite{SM_Ravi} and described many body wave function for arbitrary polarized QH states. The most successful theory to explain the partially polarized states of FQHE is the CF theory\cite{Park_Jain}. 
 
\begin{figure}
 \centering
   \includegraphics[width=8cm]{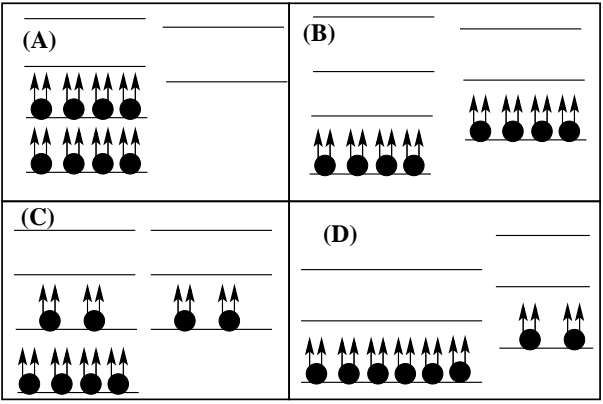}
   \caption{In each block left panel represents up-spin $\Lambda$-level, and right panel represents the down-spin $\Lambda$-level. Solid dot with arrow lines represents a CF, electron (solid dot) with even number of magnetic flux attachment(arrow lines). The FQHE filling fraction of electron $\nu=2/5(2/3)$ maps into $n=2$ filled $\Lambda$ level (CF Landau level). (A) Fully polarized state, two spin-up $\Lambda$-levels are filled. (B) Unpolarized state, one spin-up $\Lambda$-level filled  and one spin-down $\lambda$-level filled.  (C) Partially polarized Murthy's density wave state one spin-up $\Lambda$-level filled and one spin-up $\Lambda$-level + one spin-down $\lambda$-level half filled.  (D) partially polarized state, one spin-up $\Lambda$-level filled  and one spin-down $\lambda$-level filled, but the density of states for the up-spin $\Lambda$-level and spin-down $\lambda$-level are different in such a way that $\gamma = 1/2$. }
   \label{Pol-diagram}
 \end{figure}

In CF picture, the FQHE of filling fraction $\nu = \frac{n}{2pn\pm 1}$ maps into non-interacting $n$ (integer)-number of filled CF $\Lambda$-levels, out of that $n_\uparrow ( n_\downarrow )$ be the number of occupied spin-up (spin-down) CF Landau bands, then the total number of filled $\Lambda$-levels $ n = n_\uparrow  + n_\downarrow $, so that the measure of polarization of the state will be $\gamma  = \frac{n_\uparrow  - n_\downarrow}{n_\uparrow  + n_\downarrow} $\cite{Park_Jain}. In this picture we have some limited number of precisely defined polarized states. In FIG. \ref{Pol-diagram}  we have explicitly explained the CF polarized states for the filling fraction $\nu=2/3$ and 2/5. Panel (A) of this figure represents the fully polarized state, in which only spin-up $\Lambda$ levels are occupied, unpolarized state is represented in the panel (B), where one spin-up and one spin-down $\Lambda$ level filled.

Kukushkin, Klitzing and others\cite{Kukushkin} measured magnetic field dependencies of the electron spin polarization for various filling fractions ($ \nu=2/3,3/5,4/7,2/5,3/7,4/9) $. Beside the CF polarized states, they observed some specific polarized states which are not explained by the CF picture. The partially polarized states of 2/5 filling fraction has been addressed by Ganpathy Murthy\cite{Murthy 2000} as Hofstadter butterfly problem of charge density wave states of partially filled CF Landau levels (panel (C) of FIG. \ref{Pol-diagram}).
There is an exact-diagonalization calculation\cite{Wojs07, paired} for limited number of particles on sphere and torus geometry suggesting anti-ferromagnetic ordered states of $\gamma=1/2$ at 2/5 and 2/3 filling fractions, also they have demanded a calculation on the large number of particles to understand the occurrence and incompressibility of states.  
The multi-flavoured CF picture, $\Lambda$ levels within $\Lambda$ level, has been applied on the partially poarized Hall states outside Jain series\cite{CF_multi_flavor}, though we do not have any clear understanding of attaching different number of flux quanta with the same kind of electrons.  
 In this article we have tried to explain partially polarized states using Chern-Simon's (CS) theory and Plasma (PL) picture.
   
\section{Composite Fermion-Chern Simon's theory}
 
CF is a topological quantity since, they attach quantized vortices with them. The vortices bound to them produce Berry phases, that partly cancel the Aharonov-Bohm phase due to the external magnetic field. By an unitary transformation, so called CS transformation \cite{Lopez_Fradkin} to the electron field operators leads to a topological vector field $ \overrightarrow{a}_\alpha$, here $\alpha$ represent spin indices ($\uparrow  \; and \downarrow $). The corresponding CS field is given by,
 \begin{equation}
  b_\alpha = 1/e \; \overrightarrow{\nabla} \times \overrightarrow{a}_\alpha  = \phi _0 \hspace{1ex} K_{\alpha \beta } \hspace{1ex} \rho _\beta
  \end{equation}
  here, $K_{\alpha \beta}$ is the two dimensional coupling matrix, $\rho_\beta$ is the density of electron in the $\beta$ spin segment, here summation convention has been used.
Each species of CS-CF (the quasi-particle under CS transformation) will experience different effective magnetic field. The relation between these mean effective fields and the total applied physical field B is given by
  \begin{equation}
  B^*_\alpha = B - \phi _{0} \hspace{1ex} K_{\alpha \beta } \hspace{1ex} \rho _{\beta }
  \label{effB}
  \end{equation}
 In the magnetic flux attachment picture, we can explain this as an up-spin electron captures $K_{11}$ flux quantum of magnetic field, whereas a down-spin electron captures $K_{22}$ number of flux quantum of magnetic field, and an up-spin electron feels that a down-spin electron is attached with $K_{12}$ number of flux quantum of magnetic field. So $K_{11}$ and $K_{22}$ must be even integers, otherwise we shall lose the Fermionic nature. A special case if we set all the elements equal i.e. $K_{11}=K_{22}=K_{12}=K_{21}$, the state becomes Jain's CF state. 
 
The spin-dependent effective magnetic field creates different set of effective Landau levels with different degeneracy in different spin segment (panel (D) of FIG. \ref{Pol-diagram}) as degeneracy is proportional to the magnetic filed. Denoting $ n _ \uparrow ,\; n _\downarrow $ as the number of completely filled effective Landau levels ($\Lambda$ levels)  by the spin-up and spin-down species of CS-CFs, one obtains the relation from equation (2),
  \begin{eqnarray}
    \frac{\rho _{\alpha }}{n _{\alpha }} = \frac{\rho }{\nu } - K_{\alpha \beta } \hspace{1ex}\rho _{\beta }
  \label{coupled_equn}
  \end{eqnarray}
  The polarization is $ \gamma  = \frac{(\rho _{\uparrow}-\rho _{\downarrow})}{\rho } $ and total density is $ \rho =\rho _{\uparrow} + \rho _{\downarrow}$

We have studied Chern-Simon's wave function $ (k_{1},k_{2},n) $ for the SU(2) case, which is described by the exponent matrix\cite{SM_KS}
 
 \[
   K_{SU(2)} =
  \left[ {\begin{array}{cc}
   2k_{1} & n \\
   n & 2k_{2} \\
  \end{array} } \right]
\]
where k’s and n’s are positive integers.
   
\subparagraph{}
Eliminating all $ \rho  's $ from equation (\ref{coupled_equn}), we find out relation between $ k_{1}, k_{2}, n $ with polarization and total filling factor $\nu $ given below,
\begin{eqnarray}
  \frac{1+\gamma }{n _\uparrow} = \frac{2}{\nu } - 2k_{1}(1+\gamma ) -n(1-\gamma ) \\
 \frac{1-\gamma }{ n _ \downarrow} = \frac{2}{\nu } - 2k_{2}(1-\gamma ) -n(1+\gamma ) 
\end{eqnarray}
The different combination of the parameters ($k_1$, $k_2$, $n$) and number of filled $\Lambda$ levels ($n_\uparrow$ and $n_\downarrow$) give us different polarized states of a particular filling fraction $\nu$.

\subparagraph{Wave Function:}
 We have considered $N$ number of electrons moving on a spherical surface \cite{Jain_book, Fano's} in presence of radial magnetic field of total flux $2Q\phi_0$ created by a Dirac monopole at the center of sphere with monopole strength $Q$. The radius of the sphere is $R = \sqrt{Q} $, in units of magnetic length $l=\sqrt{\hbar c/eB }$.
The effective magnetic flux experience by the composite particles can be expressed from equation (\ref{effB}) as
\begin{equation}
  2q_\alpha = 2Q - \sum _{\beta }(N_\beta-\delta_{\alpha \beta}) K_{\alpha \beta}
\end{equation}

\begin{eqnarray}
  \Rightarrow Q = \frac{1}{4} \left (\sum 2q_\alpha + \sum 2 k_\alpha (N_\alpha-1) + n N\right ) \nonumber \\ 
        = \frac{1}{4} ( 2q_1 +2q_2 +  2 k_1 (N_1-1) + 2 k_2 (N_2-1) + n N ) \nonumber
\end{eqnarray}

The above relation is very important to set the radius of the spherical surface. Here, $ N_1 (N_2) $ is the number of spin up (down) electrons. The variational wave function for the state is proposed by S. Mandal and coworker\cite{SM_KS} and is given by

\begin{eqnarray}
\Psi _{k_1, k_2, n}  &=&  {\cal P}_L \Phi_{n_\uparrow}(\Omega_1^{(1)},\cdots,\Omega_{N_1}^{(1)}) \Phi_{n_\downarrow}(\Omega_1^{(2)},\cdots,\Omega_{N_2}^{(2)})  \nonumber \\
&&\times J_{11} \;J_{22}\;  J_{12}  \nonumber 
\label{Wavefn}
\end{eqnarray}

where $\Phi_{n_\uparrow}$ is the Slater determinant of $n_\uparrow$ filled CFs $\Lambda$ level , $\Omega_1^{(i)}, \cdots \Omega_N^{(i)}$ are the positions of CFs on the spherical surface, upper index indicate the different species of CFs and the Jastrow factor is given by,
\begin{eqnarray}
  J_{1 2} &=&  \prod_{i, j}^{N_1, N_2} (u_i^{(1)} v_j^{(2)}-u_j^{(2)} v_i^{(1)})^{n} \nonumber \\
  J_{\alpha \alpha}& =&  \prod_{i< j}^{N_\alpha} (u_i^{(\alpha)} v_j^{(\alpha)}-u_j^{(\alpha)} v_i^{(\alpha)})^{2k_\alpha} \nonumber 
  \end{eqnarray}
 
  where the spinor variables are $u_{i}=cos(\theta_{i}/2) exp(-i\phi_{i} /2)$ and $v_{i}=sin(\theta_{i}/2) exp(i\phi_{i}/2)$ with $0\le \theta_{i} \le \pi$ and $0\le \phi_{i} \le  2\pi$. ${\cal P}_L$ is the lowest Landau level projection operator\cite{CF_proj, DM_SSM}.

  The ground state energy of N number of particles in this configuration is given by
    \begin{eqnarray}
      E_g = \frac{<\Psi _{k_1, k_2, n}| V |\Psi _{k_1, k_2, n}>}{<\Psi _{k_1, k_2, n}|\Psi _{k_1, k_2, n}>}-\frac{N^2}{2R}
    \end{eqnarray}
  Here, $ V =\sum_{i<j} \frac{e^2}{ r_{i,j}} $ is the coulomb interaction, where $ r_{i,j} $ is the inter-electronic distance. Second term is the interaction of electrons with uniform positive background. We have used quantum Monte Carlo method to evaluate the many body integration. We have multiplied an additional factor $  \sqrt{ 2Q \nu/N}$ to the energy to compensate finite size effect.

\section{Plasma Picture}

According to Halperin's formalism \cite{Halperin}, the wave function for two-component QH-system considering spin degrees of freedom can be represented by Laughline type wave function of two component plasma(PL) in two dimension as
\begin{equation}
\Psi _{m_1,m_2,n} = \phi _{m_1}^L \phi_{m_2}^L  \; \phi _{n}^{inter}   \mbox{~~exp} \left (-\sum_{k=1,2}\sum_{i=1}^{N_k}  \frac{\mid  z^{(k)}_i \mid  ^2}{4} \right ) \nonumber
\end{equation}
where up-spin electron and down-spin electron Laughline wave functions are 
\begin{equation}
\phi _{m_1}^{L} = \prod _{i< j}^{N_1} (z^{(1)}_{i}-z^{(1)}_{j})^{m_1}  \;\; \& \;\; \phi _{m_2}^{L} = \prod _{i< j}^{N_2} (z^{(2)}_{i}-z^{(2)}_{j})^{m_2}\nonumber
\end{equation}

\begin{equation}
\phi ^{inter}_{n}= \prod _{i=1}^{N_1} \prod _{j=1}^{N_2} (z^{(1)}_i-z^{(2)}_j)^{n} \nonumber
\end{equation}

 \subparagraph{}
 Halperin's wave function $ (m_{1},m_{2},n) $ for the SU(2) case is described by the exponent matrix \cite{plasma}
 
 \[
   M=
  \left[ {\begin{array}{cc}
   m_{1} & n \\
   n & m_{2} \\
  \end{array} } \right]
\]
 here all the exponents are positive integer numbers.

With the help of the symmetric $ K \times K $ exponent matrix $ M_k \equiv (n_{ij}) $ , where $ n_{ij} = n_{ji} $ and $ n_{jj} = m_{j} $ , one may thus calculate the component filling factor in a concise manner, as stated by Goerbig and Regnault\cite{Goerbig}
\begin{equation}
\nu = M^{-1} \textbf{I} ;   
\end{equation}
 \textbf{I}  is column matrix of order 2 with elements identity.
 \[
 \left[ {\begin{array}{cc}
   \nu _{1} \\
   \nu _{2} \\
  \end{array} } \right] = M^{-1} \left[ {\begin{array}{cc}
   1 \\
   1 \\
  \end{array} } \right]
\]
Then we have,\begin{equation}
\nu _{1}= \frac{m_{2}-n}{m_{1}m_{2}-n^{2}}  ; \hspace{2ex} \nu _{2}= \frac{m_{1}-n}{m_{1}m_{2}-n^{2}}  
\end{equation}
\begin{equation}
\nu _{T}= \nu _{1} + \nu _{2} = \frac{m_{1}+m_{2}-2n}{m_{1}m_{2}-n^{2}}
\end{equation}

Positive densities (filling factors) are found only for 
\begin{equation}
m_{1}\geq{n}  \hspace{2ex}  and \hspace{2ex} m_{2}\geq{n} \nonumber
\end{equation}

Then polarization is given by,

\begin{eqnarray}
\gamma  = \frac{\nu _{1}-\nu _{2}}{\nu _{1}+ \nu _{2}} = \frac{m_{2}-m_{1}}{m_{1}+m_{2}-2n}
\end{eqnarray}

The state (m, m, n) i.e. $ m_{1}=m_{2}=m $ ; 
\begin{eqnarray}
\nu _{1}=\nu _{2}= \frac{1}{m+n};  \hspace{2ex}
\nu _{T}= \frac{2}{m+n}
\end{eqnarray}
 gives only $\gamma = 0 $ 


\begin{figure}
\centering  
\includegraphics[width=8.5cm]{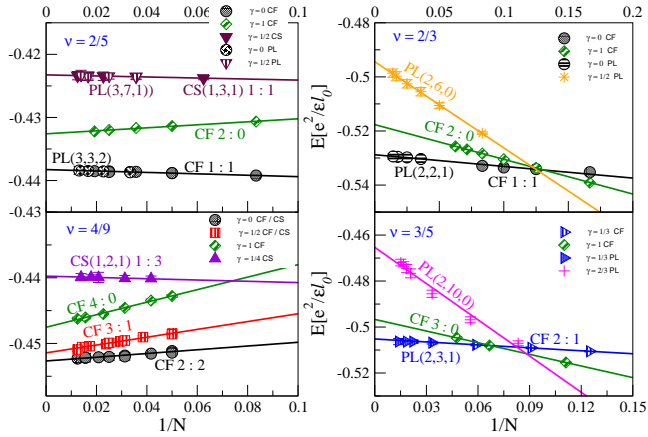}
\caption{(Color online) Estimation of ground state energy per particles. We have calculated the ground state energy for finite number of particles, to get the thermodynamic limit we have taken the extrapolation of energy in the $1/N \; \rightarrow 0$ limit, $N$ is the number of particles. Energy has been expressed in natural unit $e^2/\epsilon l$, $\epsilon$ is the dielectric constant of the background material. Result of CF theory have been shown to compare the results of our calculation. Error of the Monte Carlo integration is less than the symbol size. The states $PL(m_1,m_2,n) $ and $ CS(k_1,k_2,n) n_ \uparrow : n _\downarrow $ are shown close to the data points, which are also listed in the following table.}
\end{figure}

 \begin{figure}
\centering  
\includegraphics[width=8cm]{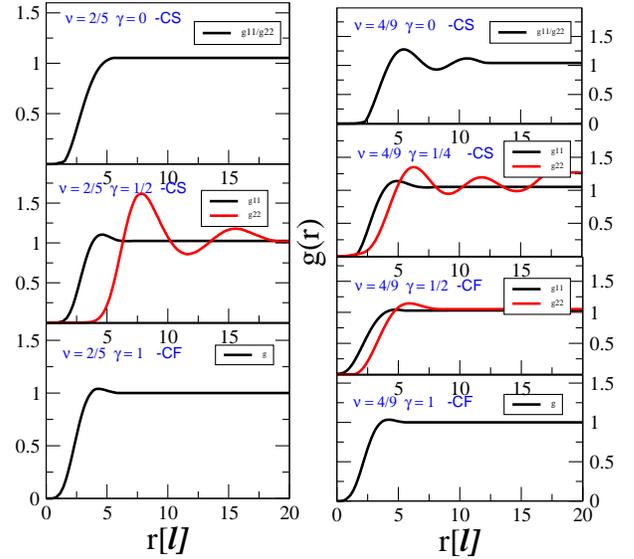}
\caption{(Color online) Pair correlation function g(r) of different polarized state for $ \nu = 2/5,4/9 $.  N=100 and more number of electrons has been used to compute g(r), for CS and CF wave function.}
\end{figure}

 \begin{table}
   \caption{Ground state energy per particles of different states}
   \label{tab:}
 
   \begin{center}
   CF-CS calculation
     \begin{tabular}{|c|c|c|c|c|c|c|c|c|}\hline
$\nu$& $\gamma $& $n_\uparrow$& $n_\downarrow$ & $k_1$ &$k_2$ & $ n $ & $E_g(CS)$ \\ 
\hline \hline
       
       2/5  & 0 & 1 & 1 & 1 & 1 & 2 & -0.438266  \\  \hline
       2/5  & 1/2 & 1 & 1 & 1 & 3 & 1 & -0.42326   \\ \hline
       2/5  & 1 & 2 & 0 & 1 & - & - &  -0.432597  \\   \hline \hline

       4/9 & 0 & 2 & 2 & 1 & 1 & 2 & -0.45259  \\  \hline
       4/9  & 1/4 & 1 & 3 & 1 & 2 & 1 & -0.43975   \\ \hline
       4/9 & 1/2 & 3 & 1 & 1 & 1 & 2 &  -0.45158 \\  \hline
       4/9 & 1 & 4 & 0 & 1 & - & - &  -0.447535 \\   \hline 
       
      \end{tabular} \\
      CF and PL calculation
      \begin{tabular}{|c|c|c|c|c|c|c|c|c|}\hline
 $\nu$& $\gamma $& $\nu_1$& $\nu_2$ & $m_1$ &$m_2$ & $ n $ & $ E_g(PL)$ &$E_g(CF)$\\ \hline

       2/5  & 0 & 1/5 & 1/5 & 3 & 3 & 2 & -0.438266 & -0.438266  \\  \hline
       2/5  & 1/2  & 3/10 & 1/10 & 3 & 7 & 1 & -0.42332 & - \\ \hline \hline
        
       2/3 & 0 & 1/3 & 1/3 & 2 & 2 & 1 &  -0.52879 & -0.52929 \\ \hline
       2/3 & 1/2 & 1/2 & 1/6 & 2 & 6 & 0  & -0.49438 & -  \\ \hline
       2/3 & 1 & 2 & 0 & - & - & -  & - & -0.51906 \\ \hline \hline
       
       3/5 & 1/3 & 2/5 & 1/5 & 2 & 3 & 1  & -0.50588 & -0.5052 \\ \hline
       3/5 & 2/3 & 1/2 & 1/10 & 2 & 10 & 0 &  -0.46534 & - \\ \hline
       3/5 & 1 & 3 & 0 & - & - & - &  - & -0.49668 \\ \hline        
     \end{tabular}
   \end{center}
 \end{table}

\section{Pair correlation function}

An important quantity for a liquid is its pair distribution function $g(r)$. The liquid phase is defined by the property that, $g(r)$ approaches a constant at large $r$. The pair distribution function $g(r)$ is the probability of finding two particles at a distance $r$ ($r = |\vec{r}_1-\vec{r}_2$|) 

\begin{eqnarray}
  g^{\alpha \alpha } (r) = \frac{N(N-1)}{\rho } \int d^{2}r_{3}d^{2}r_{4} \cdots d^{2}r_{N}  \\ 
      \mid\Psi (r_{1}, r_{2}, \cdots,r_{N})\mid^{2}  \nonumber
\end{eqnarray} 

$\Psi$ is the many body wave function,    $ g^{\alpha \alpha } $ denotes probability of finding two particles at a distance r in the same spin segment $\alpha$, ($\vec{r}_1$ and $\vec{r}_2$ are in the spin segment $\alpha$).

The pair correlation function of a FQHE state can be obtained with the help of the ground state wave function, where multidimensional integrals are evaluated numerically by Monte Carlo methods. The systems are sufficiently large that the results are essentially independent of particle numbers, and can be considered to represent the thermodynamic limit. Generally, g(r) becomes constant for large r, confirming that the wave functions describe a liquid. For a crystal with long-range order, the pair correlation function would oscillate all the way to infinity.
At very low filling fraction of a LL, fractional quantum Hall state (FQHS) is destroyed and form Wigner crystal state\cite{WC, bubbles, strips}, either in bubble form or in strip form.

\begin{figure}
\centering  
\includegraphics[width=8cm]{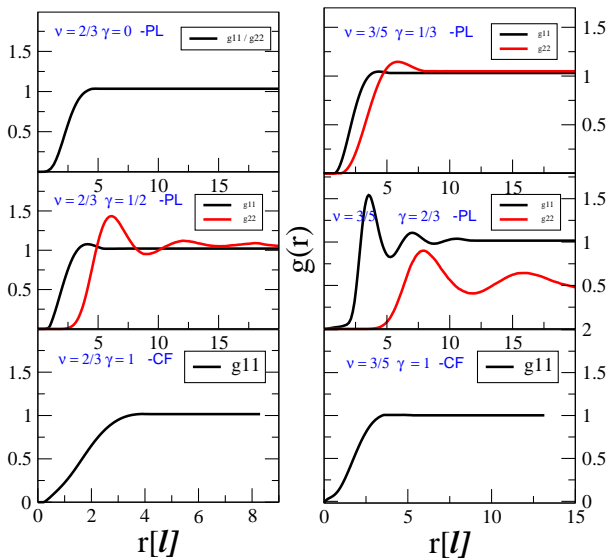}
\caption{(Color online) Pair correlation function g(r) of different polarized state for $ \nu = 2/3, 3/5 $ at large number of electrons is computed using parallel program (MPI) for CF and PL wave function. It is not possible to perform Monte Carlo calculation for large number of particles for 2/3 and 3/5 filling fraction in the CF picture due to complicated projected wave function on the lowest LL, that is why we have shown results for small number of electrons( upto 14 electrons for 2/3 filling fraction, and 21 electrons for 3/5 filling fraction) in the lowest panel. }
\end{figure}

 \section{Results \& Discussion}
 
 Partially polarized ($\gamma = 1/2$) state of 2/5 filling fraction has been studied using CF-CS theory as well as plasma picture. Both the calculation give the same energy, but the energy of the state is higher than the fully polarized state, which is unexpected. The pair correlation function calculation shows that the system is not an ideal liquid state, up-spin electrons behave as liquid state whereas, down-spin electrons behave more like crystalline state. So we predict that, the partially polarized state of 2/5 consists of two phases of up-electron liquid phase and down-electron solid phase. Our prediction slightly differs from Murthy's prediction. Murthy proposed that the partially polarized 2/5 state is a Hofstadter's butterfly charge density wave state with partially filled $\Lambda$ levels form a CF-crystal. 
 
Beside 2/5 filling fraction CF-CS theory fit with $\gamma=1/4$ polarized state of 4/9 filling fraction. In this case the energy again becomes higher than the fully polarized state. The pair correlation function calculation shows that the state consists of up-electron liquid state and down-electron crystalline state.
 In addition of CF-CS theory we have investigated the Halperin's states in which up-spin and down-spin LL are partially filled. In our study we have seen a very interesting result that both the CF theory and PL give the exactly same energy of the state $\gamma=1/3$ at $\nu=3/5$ filling fraction, though the two formalism are different. The state $\gamma=2/3$ at $\nu=3/5$ and $\gamma = 1/2$ at $\nu=2/3$ fit with PL with parameter (2,10,0) and (2,6,0) respectively. Both the states have higher energy than that of fully polarized states. Pair correlation calculation shows that both the states are consist of liquid state of spin-up electrons and crystalline state of spin-down electrons, similar to the CF-CS study of $\nu=2/5$ and $\nu=4/9$.  
 The partially polarized states out of Jain's main sequence, have higher energy than the fully polarized states, apparently contradicts the experimentally observed states. We believe that the electrons in the liquid states take part in the inelastic Raman scattering experiment, may be the contribution to the energy of the electrons in solid state is much more than liquid state, that's why the average energy per particle is higher than that of fully polarized state. 
 So we predict that the observed polarized states with small plateau in the polarized diagram\cite{Kukushkin} must be consisted of the liquid phase of up-spin electrons and crystalline phase of down-spin electrons.

\section{acknowledgement}

We would like to thank Sudhansu S. Mandal for fruitful discussions. MI acknowledges DST INSPIRE and DD thanks UGC, India for the financial support.


\begin{thebibliography}{}

  \bibitem{fqhe} D. C. Tsui, H. L. Stormer and A. C. Gossard, Phys. Rev. Lett. {\bf 48}, 1559 (1982), Phys. Rev. B {\bf 25}, 1405 (1982) ; H. L. Stormer, A. Chang, D. C. Tsui, J. C. M. Hwang, A. C. Gossard and W. Wiegmann, Phys. Rev. Lett. {\bf 50}, 1953 (1983).
  \bibitem{Jain_CF} J. K. Jain, Phys. Rev. Lett. {\bf 63}, 199 (1989), Phys. Rev. B {\bf 41}, 7653 (1990).
   \bibitem{Jain_CF1} A. S. Goldhaber and J. K. jain, Physics Letters A {\bf 199} 267 (1995).
  \bibitem{DDMI} D. Das, M. Indra and D. Majumder, Solid State Communications {\bf 260} (2017) 19–22.
 \bibitem{DM_SSM} D. Majumder and S. S. Mandal, Phys. Rev. B {\bf 90}, 155310 (2014).
 
 \bibitem{a} R. G. Clark, S. R. Haynes, A. M. Suckling, J. R. Mallett, P. W. Wright, J. J. Harris and C. T. Foxon, Phys. Rev. Lett. {\bf 62}, 1536 (1989).
  \bibitem{b} J. P. Eisenstein, H. L. Stormer, L. N. Pfeiffer and K. W. West, Phys. Rev. Lett. {\bf 62}, 1540 (1989).
  \bibitem{c} J. P. Eisenstein, H. L. Stormer, L. N. Pfeiffer and K. W. West, Phys. Rev. B {\bf 41}, 7910 (1990).       
  \bibitem{d} L. W. Engel, S. W. Hwang, T. Sajoto, D. C. Tsui and M. Shayegan, Phys. Rev. B {\bf 45}, 3418 (1992).
 
 \bibitem{LLMixing} Y. Zhang, A. Wojs and J. K. Jain, Phys. Rev. Lett. {\bf 117}, 116803 (2016).
 \bibitem{exact_dia} Song He, S. H. Simon and B. I. Halperin, Phys. Rev. B {\bf 50}, 1823 (1994);  F. D. M. Haldane and
E. H. Rezayi, Phys. Rev. Lett. {\bf 54}, 237 (1985). X. C. Xie, Y. Guo, and F. C. Zhang, Phys. Rev. B {\bf 40}, 3487 (1989); D. Yoshioka, J. Phys. Soc. Jpn. {\bf 55}, 885 (1986); F. C. Zhang and T. Chakraborty, Phys. Rev. B {\bf 30}, 7320 (1984).
  \bibitem{extrapolation} K. Park and J. K. Jain, Phys. Rev. Lett. {\bf 83}, 5543 (1999) ; A. C. Balram, C. Toke, A. Wojs and J. K. Jain, Phys. Rev. B {\bf 92}, 075410 (2015).
  
  \bibitem{SM_Ravi} S. S. Mandal  and V. Ravishankar, Phys. Rev. B {\bf 54}, 8688 (1996).
  \bibitem{Park_Jain} K. Park and J. K. Jain, Solid State Communications {\bf 119} (2001) 291.
  
  \bibitem{Kukushkin} I. V. Kukushkin, K. v. Klitzing and K. Eber, Phys. Rev. Lett. {\bf 82}, 3665 (1999).
  \bibitem{Murthy 2000} G. Murthy, Phys. Rev. Lett. {\bf 84}, 350 (2000).
     
  \bibitem{Wojs07}K. Vyborny,O. Certik, D. Pfannkuche, D. Wodzinski, A. Wojs, and J. J. Quinn, Phys. Rev. B {\bf 75}, 045434 (2007).
   \bibitem{paired} A. Wojs, K. S. Yi and J. J. Quinn Phys. Rev. B {\bf 69}, 205322 (2004).
  \bibitem{CF_multi_flavor} A. C. Balram, C. Toke,  A. Wojs and J. K. Jain, Phys. Rev. B {\bf 91}, 045109 (2015); A. Wojs, G. Simion and J. J. Quinn hys. Rev. B {\bf 75}, 155318 (2007).
 
  \bibitem{Lopez_Fradkin} A. Lopez and E. Fradkin, Phys. Rev. B {\bf 44}, 5246 (1991).
  \bibitem{SM_KS} S. Modak, S. S. Mandal and K. Sengupta, Phys. Rev. B {\bf84}, 165118 (2011).
  
  \bibitem{Jain_book} Composite Fermions, J. K. Jain ( Cambridge University Press), http://www.cambridge.org/9780521862325
  \bibitem{Fano's} G. Fano, F. Ortolani and E. Colombo, Phys. Rev. B {\bf 34}, 2670 (1986).
  
  \bibitem{CF_proj} J. K. Jain and  R. K. Kamilla, Phys. Rev. B {\bf 55}, R4895(1997);  Int. J. Mod. Phys. B {\bf 11}, 2621
(1997).
   	 
  

  \bibitem{Halperin} B. I. Halperin, Helv. Phys. Acta {\bf 56}, 75 (1983).
  \bibitem{plasma} R. de Gail, N. Regnault and M. O. Goerbig, Phys. Rev. B {\bf 77}, 165310 (2008).
  \bibitem{Goerbig} M. O. Goerbig and N. Regnault , Phys. Rev. B {\bf 75}, 241405 (2007).
  
  
  
  
  \bibitem{WC} M. R. Peterson and J. K. Jain,  Phys. Rev. B {\bf 68}, 195310 (2003); G. Gervais, L. W. Engel, H. L. Stormer, D. C. Tsui, K. W. Baldwin, K. W. West, L. N. Pfeiffer,  Phys. Rev. Lett. {\bf 93},266804 (2004).
 	\bibitem{bubbles} S. Y. Lee, V. W. Scarola and J. K. Jain, Phys. Rev. B {\bf 66}, 085336 (2002).
  \bibitem{strips} W. Pan, H. L. Stormer, D. C. Tsui, L. N. Pfeiffer, K. W. Baldwin and K. W. West, Phys. Rev. Lett. {\bf 88}, 176802 (2002).
 	 
  
  

  
  
  
  
\end{thebibliography}
\end{document}